\documentstyle[prl,eqsecnum,aps]{revtex}
\twocolumn

\begin{document}
\author{Jian-Qi Shen \footnote{E-mail address: jqshen@coer.zju.edu.cn}}
\address{Zhejiang Institute of Modern Physics and Department of Physics,
Zhejiang University, Hangzhou 310027, P.R. China}
\date{\today }
\title{Hyperbolical geometric quantum phase and topological dual mass } \maketitle

\begin{abstract}
In this note we show the existence of the hyperbolical geometric
quantum phase that is different from the ordinary trigonometric
geometric quantum phase. Gravitomagnetic charge (dual mass) is the
gravitational analogue of magnetic monopole in Electrodynamics;
but, as will be shown here, it possesses more interesting and
significant features, {\it e.g.}, it may constitute the dual
matter that has different gravitational properties compared with
mass. In order to describe the space-time curvature due to the
topological dual mass, we construct the dual Einstein's tensor.
Further investigation shows that gravitomagnetic potentials caused
by dual mass are respectively analogous to the trigonometric and
hyperbolic geometric phase. The study of the geometric phase and
dual mass provides a valuable insight into the time evolution of
quantum systems and the topological properties in General
Relativity.


\end{abstract}

Both geometric phase\cite{1} of wave function in Quantum Mechanics
and gravitomagnetic charge (topological dual charge of mass) in
the general theory of relativity reveal Nature's geometric or
global properties. Differing from the dynamical phase, geometric
phase depends only on the geometric nature of the pathway along
which the quantum system evolves\cite{3}. Geometric phase exists
in time-dependent quantum systems or systems whose Hamiltonian
possesses evolution parameters\cite{7}. As is well known, the
dynamical phase of wave function in Quantum Mechanics is dependent
on dynamical quantities such as energy, frequency, coupling
coefficients and velocity of a particle or a quantum system, while
the geometric phase is immediately independent of these physical
quantities. When Berry found that the wave function would give
rise to a non-integral phase (Berry's phase) in quantum adiabatic
process\cite{1}, geometric phase problems attract attention of
many physicists in various fields such as gravity theory\cite{9},
differential geometry\cite{3}, atomic and molecular
physics\cite{11}, nuclear physics\cite{11}, quantum
optics\cite{13}, condensed matter physics\cite{16} and molecular
reaction (molecular chemistry)\cite{11} as well. In many simple
quantum systems such as an electron possessing intrinsic magnetic
moment interacting with a time-dependent magnetic field (or a
neutron spin interacting with the Earth's rotation\cite{20}), a
photon propagating inside the curved optical fiber\cite{5}, and
the time-dependent Jaynes-Cummings model describing the
interaction of the two-level atom with a radiation field\cite{21},
geometric phase is often proportional to $2\pi(1-\cos\theta)$,
which equals the solid angle subtended by the curve with respect
to the origin of parameter space\footnote{The solid angle
subtended by a curve shows the topological meanings of geometric
phase. Geometric phase is absent in the quantum system when its
Hamiltonian is independent of time. Geometric phase of many simple
physical systems in the adiabatic process can be expressed in
terms of the solid angle over the parameter space, {\it e.g.}, the
wave function of a photon propagating inside the non-coplanar
curved optical fiber obtains this topological phase, where the
parameter space is just the momentum ({\it i.e.}, velocity) space
of the photon.}. This, therefore, implies that geometric phase
differs from dynamical phase and it involves global and
topological information on the time evolution of quantum systems.
In addition to this trigonometric geometric phase, there exists
the so-called hyperbolical geometric phase that is expressed by
$2\pi(1-\cosh\theta)$ with the hyperbolical cosine
$\cosh\theta=\frac{1}{2}\left[\exp(\theta)+\exp(-\theta)\right]$
in some time-dependent quantum systems, {\it e.g.}, the two-level
atomic system with electric dipole-dipole interaction and the
harmonic-oscillator system\cite{22}. It is verified that the
generators of the Hamiltonians of these quantum systems form the
$SU(1,1)$ Lie algebra. Further analysis indicates that quantum
systems, which possess the non-compact Lie algebraic structure
(whose group parameters can be taken to be infinity) will present
the hyperbolical geometric phase, while quantum systems with
compact Lie algebraic structure will give rise to the
trigonometric geometric phase. Since Lorentz group (describing the
boosts of reference frames in the space direction) in the special
theory of relativity is also a non-compact group, this leads us to
consider the topological properties associated with space-time. We
take into account the gravitational analogue of magnetic
charge\cite{6}, {\it i.e.}, gravitomagnetic charge that is the
source of gravitomagnetic field just as the case that mass
(gravitoelectric charge) is the source of gravitoelectric field
({\it i.e.}, Newtonian gravitational field in the sense of
weak-field approximation). In this sense, gravitomagnetic charge
is also termed dual mass. It should be noted that the concept of
the ordinary mass is of no physical significance for the
gravitomagnetic charge; it is of interest to investigate the
relativistic dynamics and gravitational effects as well as
geometric properties of this topological dual mass (should such
exist). From the point of view of differential geometry, matter
may be classified into two categories: gravitoelectric matter and
gravitomagnetic matter. The former category possesses mass and
constitutes the familiar physical world, while the latter
possesses dual mass that would cause the non-analytical property
of space-time metric. Einstein's field equation of gravitation in
general theory of relativity governs the couplings of
gravitoelectric matter (which possesses mass) to gravity
(space-time); accordingly, we should have a field equation
governing the interaction of dual matter with gravity. By making
use of the variational principle, the gravitational field equation
of gravitomagnetic matter can be obtained where the dual
Einstein's tensor is denoted by
$\frac{1}{2}\left(\epsilon_{\mu}^{\ \ \lambda\sigma\tau}{\mathcal
R}_{\lambda\sigma\tau\nu}-\epsilon_{\nu}^{\ \
\lambda\sigma\tau}{\mathcal R}_{\lambda\sigma\tau\mu}\right)$ with
$\epsilon_{\mu}^{\ \ \lambda\sigma\tau}$ and ${\mathcal
R}_{\lambda\sigma\tau\nu}$ being the four-dimensional Levi-Civita
completely antisymmetric tensor and the Riemann curvature tensor
that describes the space-time curvature, respectively\cite{Shen}.
By exactly solving this field equation, one can show that the
topological property of the solution $g_{t\varphi}(r,\theta)$ can
be illustrated as follows: solid angle subtended by the curved
${\rm C}$ showing the topological property of the gravitomagnetic
vector potential of a static gravitomagnetic charge at the origin
of the spherical coordinate system. Take the gravitomagnetic
vector potentials ${\bf g}=\left(0, 0,
\frac{2\mu}{4\pi}\cdot\frac{1-\cos\theta}{r\sin\theta}\right)$ in
spherical coordinate system, then the loop integral, $\oint_{\rm C
}{\bf g}\cdot{\rm d}{\bf l}=\mu(1-\cos\theta)$, is proportional to
the solid angle subtended by the curved ${\rm C}$ with respect to
the origin. The same situations arise in the adiabatic quantum
geometric phase (Berry's quantum phase), which reflects the global
or topological properties of time evolution (or parameters
evolution) of quantum systems. Such property is in analogy with
that of the geometric quantum phase in the time-dependent
spin-gravity coupling ({\it i.e.}, the interaction between a
spinning particle with gravitimagnetic field\cite{20}) and other
quantum adiabatic processes\cite{9,5}. The topological properties
of gravitomagnetic charge (dual mass) may be shown in terms of the
global features of geometric quantum
phase\footnote{Gravitomagnetic moment results from the mass
current, which is also the dynamical physical quantity. Comparison
between gravitomagnetic charge and geometric phase enables to show
the topological properties of the former. The reason why the
topological property is important lies in that the global
description of the physical phenomena is essential to understand
the world. It is of interest that dual matter may constitute a
dual world where dual mass abides by their own dynamical and
gravitational laws, which is somewhat different from the laws in
our world; for example, dual mass is acted upon by a
gravitomagnetic Lorentz force in Newton's gravitational
(gravitoelectric) field, and the static dual mass produces the
gravitomagnetic field rather than the Newton's gravitoelectric
field.}. It follows that the expression of gravitomagnetic
potential, $g_{t\varphi}(r,\theta)$, due to dual mass is exactly
analogous to that of the trigonometric geometric phase. In the
similar fashion, it is readily verified that the gravitomagnetic
potential, $g_{\theta\varphi}(r,t)$, is similar to that of the
hyperbolical geometric phase. This feature originates from the
fact that the Lorentz group is a non-compact group. Although there
is no evidence for the existence of this topological dual mass at
present, it is still essential to consider this topological or
global phenomenon in General Relativity. It is believed that there
would exist formation (or creation) mechanism of gravitomagnetic
charge in the gravitational interaction, just as some prevalent
theories provide the theoretical mechanism of existence of
magnetic monopole in various gauge interactions\cite{23}. Magnetic
monopole in electrodynamics and gauge field theory has been
discussed and sought after for decades, and the existence of the
't Hooft-Polyakov monopole solution\cite{23} has spurred new
interest of both theorists and experimentalists\cite{23}. As the
topological gravitomagnetic charge in the curved space-time, dual
mass is believed to give rise to such interesting situation
similar to that of magnetic monopole. If it is indeed present in
universe, dual mass will also lead to significant consequences in
astrophysics and cosmology. We emphasize that although the
gravitomagnetic vector potential produced by the gravitomagnetic
charge is the classical solution to the field equation, this kind
of topological gravitomagnetic monopoles may arise not as
fundamental entities in gravity theory, {\it e.g.}, it will behave
like a topological soliton. Gravitomagnetic charge has some
interesting relativistic quantum gravitational effects\cite{20},
{\it e.g.}, the gravitational Meissner effect, which  may serve as
an interpretation of the smallness of the observed cosmological
constant. In accordance with quantum field theory, vacuum
possesses infinite zero-point energy density due to the vacuum
quantum fluctuations; whereas according to Einstein's theory of
General Relativity, infinite vacuum energy density yields the
divergent curvature of space-time, namely, the space-time of
vacuum is extremely curved. Apparently it is in contradiction with
the practical fact, since it follows from experimental
observations that the space-time of vacuum is asymptotically flat.
In the context of quantum field theory a cosmological constant
corresponds to the energy density associated with the vacuum and
then the divergent cosmological constant may result from the
infinite energy density of vacuum quantum fluctuations. However, a
diverse set of observations suggests that the universe possesses a
nonzero but very small cosmological constant\cite{28}. How can we
give a natural interpretation for the above paradox? Here,
provided that vacuum matter is perfect fluid, which leads to the
formal similarities between the weak-gravity equation in perfect
fluid and the London's electrodynamics of superconductivity, we
suggest a potential explanation by using the canceling mechanism
via gravitational Meissner effect: the gravitoelectric field
(Newtonian field of gravity) produced by the gravitoelectric
charge (mass) of the vacuum quantum fluctuations is  exactly
canceled by the gravitoelectric field due to the induced current
of the gravitomagnetic charge of the vacuum quantum fluctuations;
the gravitomagnetic field produced by the gravitomagnetic charge
(dual mass) of the vacuum quantum fluctuations is exactly canceled
by the gravitomagnetic field due to the induced current of the
gravitoelectric charge (mass current) of the vacuum quantum
fluctuations. Thus, at least in the framework of weak-field
approximation, the extreme space-time curvature of vacuum caused
by the large amount of the vacuum energy does not arise, and the
gravitational effects of cosmological constant is eliminated by
the contributions of the gravitomagnetic charge (dual mass). If
gravitational Meissner effect is of really physical significance,
then it is necessary to apply this effect to the early universe
where quantum and inflationary cosmologies dominate the evolution
of the universe. Study of the geometric property in quantum
regimes is an interesting and valuable direction. Since it reveals
the global and topological properties of evolution of quantum
systems, geometric phase has many applications in various branches
of physics, say, in the coupling of neutron spin to the Earth's
rotation\cite{20}, a potential application may be suggested where
the information on the Earth's variations of rotating frequency
will be obtained by measuring the geometric phase of the
oppositely polarized neutrons through the neutron-gravity
interferometer experiment. The topological charge in curved
space-time also deserves further investigation, since it reflects
plentiful global or geometric properties hidden in the gravity
theory. It is believed that both theoretical and experimental
interest in this direction may enable people to understand the
global phenomena of the physical world better.

\end{document}